\documentclass[twocolumn,10pt,prl,nofootinbib,preprintnumbers]{revtex4-1}

\usepackage[utf8]{inputenc}

\def\mysections#1{{\bf #1.} }
\usepackage{hyperref}

\usepackage[dvips]{graphicx}

\usepackage{epsfig,amsmath,amssymb,verbatim,mathrsfs,array,layout,textcomp,amssymb,latexsym,slashed,graphicx,booktabs,color,mathtools,tikz}

\usepackage{cleveref}

\newcommand{\beq}{\begin{equation}}
\newcommand{\eeq}{\end{equation}}
\newcommand{\bea}{\begin{eqnarray}}
\newcommand{\eea}{\end{eqnarray}}

\def\beqa{\begin{eqnarray}}
\def\eeqa{\end{eqnarray}}
\newcommand{\no}{\nonumber}
\newcommand{\bv}{\left(\begin{array}{c}}
\newcommand{\ev}{\end{array}\right)}
\newcommand{\bmtwo}{\left(\begin{array}{cc}}
\newcommand{\bmthree}{\left(\begin{array}{ccc}}
\newcommand{\emn}{\end{array}\right)}
\newcommand{\bmtwoc}{\left\{\begin{array}{cc}}
\newcommand{\bmthreec}{\left\{\begin{array}{ccc}}
\newcommand{\emnc}{\end{array}\right\}}
\newcommand{\ba}{\begin{array}}
\newcommand{\ea}{\end{array}}

\def\lsim{\mathrel{\rlap{\lower4pt\hbox{\hskip1pt$\sim$}}
     \raise1pt\hbox{$<$}}}         
\def\gsim{\mathrel{\rlap{\lower4pt\hbox{\hskip1pt$\sim$}}
     \raise1pt\hbox{$>$}}}         

\begin{document}

\font\mini=cmr10 at 0.8pt

\title{
Thermal Dark Matter from Freezeout of Inverse Decays
}

\author{Ronny Frumkin${}^{1}$}\email{ronny.frumkin@mail.huji.ac.il}
\author{Yonit Hochberg${}^{1}$}\email{yonit.hochberg@mail.huji.ac.il}
\author{Eric Kuflik${}^{1}$}\email{eric.kuflik@mail.huji.ac.il}
\author{Hitoshi Murayama${}^{2,3,4,5}$}\email{hitoshi@berkeley.edu, hitoshi.murayama@ipmu.jp}
\affiliation{${}^1$Racah Institute of Physics, Hebrew University of Jerusalem, Jerusalem 91904, Israel}
\affiliation{${}^2$Ernest Orlando Lawrence Berkeley National Laboratory, University of California, Berkeley, CA 94720, USA}
\affiliation{${}^3$Department of Physics, University of California, Berkeley, CA 94720, USA}
\affiliation{${}^4$Kavli Institute for the Physics and Mathematics of the
  Universe (WPI), University of Tokyo,
  Kashiwa 277-8583, Japan}
  \affiliation{${}^5$The Institute for AI and Beyond, The University of Tokyo, Tokyo 113-8655, Japan}

\begin{abstract}

We propose a new thermal dark matter candidate whose abundance is determined by the freezeout of inverse decays. The relic abundance  depends parametrically only on a decay width, while matching the observed value requires that the coupling determining the width---and the width itself---should be exponentially small. The dark matter is therefore very weakly coupled to the Standard Model, evading conventional searches. This INverse DecaY (`INDY') dark matter  can be discovered by searching for the long-lived particle that decays into the dark matter at future planned experiments. 
\end{abstract}

\maketitle

\section{Introduction}\label{sec:intro}

The identity of dark matter (DM) is one of the most pressing open questions in modern day physics. While the weakly interacting massive particle (WIMP) paradigm has long guided the particle physics community, the absence of experimental evidence for the WIMP at colliders, direct-detection and indirect-detection experiments stresses the importance of considering DM beyond the WIMP. Indeed, recent years have seen a surge of new DM ideas (see {\it e.g.} Refs.~\cite{Hochberg:2014dra,Hochberg:2014kqa,Griest:1990kh,DAgnolo:2015ujb,Kuflik:2015isi,Kuflik:2017iqs,Dror:2016rxc,Dror:2017gjq,Kopp:2016yji,DAgnolo:2017dbv,DAgnolo:2018wcn,Fitzpatrick:2020vba,Kim:2019udq,Asadi:2021yml,Asadi:2021pwo}) which utilize various processes in the early universe. 

 One such process is inverse decay, where a DM particle is produced through the inverse decay of a heavier particle in the dark sector. Thus far, decays have been considered in the literature in the context of freeze-in DM~\cite{Hall:2009bx}, where a slow inverse decay of a bath particle slowly freezes in the DM abundance; as a process maintaining chemical equilibrium  within the dark sector or with the SM~\cite{Dror:2016rxc,Dror:2017gjq,Berlin:2016vnh,Morrissey:2009ur,Cohen:2010kn,Bandyopadhyay:2011qm,Farina:2016llk,Kim:2019udq,Hochberg:2018vdo}; and in other dark matter frameworks~\cite{Feng:2003xh,Kaplinghat:2005sy,Moroi:1999zb,Acharya:2009zt,Hall:2009bx,Berlin:2016vnh,Morrissey:2009ur}. The effects of inverse decays on dark matter depletion have been considered as a contributing reaction~\cite{Garny:2017rxs,Yaguna:2008mi}, but never as the main process for setting the dark matter abundance. In this work, and in a companion paper~\cite{future}, we study the freezeout of inverse decays as the mechanism to set the relic abundance of DM. 
 
 This {\it Letter} is organized as follows. We begin by outlining the basic idea for freezeout of inverse decays and derive an analytical understanding of the mechanism. We then solve the Boltzmann equations of the system and obtain the DM parameter space. Finally we present a model that realizes the mechanism.

\section{Basic Idea}\label{sec:basic}
 
Here we show that the freezeout of inverse decays can be responsible for the relic abundance of DM. 
Consider a dark matter particle $\chi$ and an unstable dark sector particle $\psi$ that has a decay that contains some number of $\chi$ particles in the final state. For simplicity we will consider a simple decay and inverse decay
\beq
\psi \longleftrightarrow \chi + \phi
\eeq 
motivated by a $Z_2$ symmetry in the dark sector. (Other inverse decay topologies can be considered as well.)  Here $\phi$ can be a dark sector or visible particle that is assumed to be in equilibrium with the bath. (Later we will take a concrete model in which $\phi$ is a dark  photon that kinetically mixes with hypercharge.) 

 The Boltzmann equation for the abundance of $\chi$, assuming that $\phi$ is in equilibrium, is:
\beq
\dot{n}_\chi + 3 H n_\chi = \Gamma\left(n_\psi-n_\chi \frac{n_\psi^{\rm eq}}{n_\chi^{\rm eq}}\right)\,,
\label{eqref:bechi}
\eeq
with $\Gamma$ the decay rate of $\psi\to \chi \phi$. We assume that the DM is cold, namely that it freezes out when non-relativistic; our numerics presented later on confirm this. We can thus ignore the thermally averaged time dilation that would normally appear in the collision term. 

Approximate analytic solutions to the Boltzmann equations can be obtained in the instantaneous freezeout approximation,  but will not always suffice. The inverse decay rate is falling off exponentially as $e^{-(m_\psi-m_\chi)/T}$ (rather than $e^{-m_\chi/T}$ for the well-studied WIMP), which is not necessarily fast enough to assume  that instantaneous freezeout occurs.  Further consideration must also be taken into account because the decays and inverse decays may not actually be in equilibrium before they completely decouple.

We begin by calculating the relic abundance when decays are in equilibrium. This will give us an approximate range of parameter space---couplings and masses---necessary to reproduce the observed abundance.  We first assume that $\psi$ is always in chemical equilibrium with the Standard Model (SM) bath. This can be achieved through rapid annihilations of $\psi$ into bath particles (either SM bath particles, or particles that are in equilibrium with the SM). The Boltzmann Eq.~\eqref{eqref:bechi} can be rewritten as 
\beq
\frac{\partial Y_\chi}{\partial x}= -\frac{\Gamma}{H x}\left(Y_\chi \frac{n^{\rm eq}_\psi}{n_\chi^{\rm eq}} - Y_\psi^{\rm eq} \right)\,,
\eeq 
where $Y=n/s$ is the yield and $x=m_\chi/T$. In the instantaneous freezeout approximation, the $\chi$ particle departs the chemical equilibrium  when the coefficient of the collision term becomes of order unity,
\beq
\Gamma\frac{n^{\rm eq}_\psi}{n_\chi^{\rm eq}}\simeq xH\qquad {\rm for  }~~ x=x_f\,.
\label{eq:xf}
\eeq
The freezeout abundance of $\chi$ can be determined by solving Eq.~\eqref{eq:xf} for  $x_f$ and using that at the time of freezeout,  $\chi$ is approximately in equilibrium:
\beq
n_\chi(x_f) \simeq n_\chi^{\rm eq}(x_f) = g_\chi \left(\frac{m^2}{2\pi x_f}\right)^{3/2} e^{-x_f}\,.
\eeq

We parameterize the decay rate as   
\beq \label{eq:decay}
\Gamma \equiv \alpha_{\rm decay}m_{\psi}\sqrt{1-2\dfrac{\left(m_{\phi}^{2}+m_{\chi}^{2}\right)}{m_{\psi}^{2}}+\dfrac{\left(m_{\phi}^{2}-m_{\chi}^{2}\right)^{2}}{m_{\psi}^{4}}}\,.
\eeq
Then requiring that we match the observed  DM abundance, given by $n_\chi(x_f)\sim m^2 T_{\rm eq}$, we arrive at the requisite relationship between the DM mass and coupling,
\beq
m_\chi \simeq  \alpha_{\rm decay}^{\frac{1}{1+\Delta}}\left( m_{\rm pl} T_{\rm eq}^\Delta \right)^{\frac{1}{1+\Delta}}\,. \label{eq:mvsalpha}
\eeq
Here $\Delta \equiv (m_\psi-m_\chi)/m_\chi$, $T_{\rm eq}= 0.8$~eV is the temperature at matter radiation equality, and ${m_{\rm pl} = 2.4\times 10^{18}}$~GeV is the reduced Planck mass. This formula may remind the reader of other dark matter freezeout mechanisms, and would produce similar parametric dependence on the temperature of equality and Planck scale in the WIMP and SIMP~\cite{Hochberg:2014dra} cases  for $\Delta=1$ and $\Delta=2$, respectively. We therefore see that for small mass splittings $\Delta<1$, much heavier mass DM for the same size coupling (or much smaller coupling for same mass) is needed to produce the relic abundance, when compared to the WIMP. For this reason, inverse decays can achieve  super heavy dark matter with masses beyond the WIMP unitarity bound, in the spirit of Ref.~\cite{Kim:2019udq}, where an inverse chain of dark sector particles set the abundance~\cite{tal}. 

The above also indicates when this analysis is valid. Provided that the rate over Hubble becomes of order unity before $x_f$, the system will come close to equilibrium and there will be no sensitivity to the initial conditions. Looking at Eq.~\eqref{eq:xf}, the rate over Hubble is not monotonically decreasing, but rather reaches a  maximum at $x=\Delta^{-1}$. Equilibrium is achieved if 
\beq
\Gamma \gtrsim \Delta H_m\,, \label{eq:eqsol}
\eeq
where $H_m \equiv H(x=1)$. 
If Eq.~\eqref{eq:eqsol} is not satisfied, then the decays and inverse decays are not sufficient to thermalize the dark matter to chemical equilibrium before freezeout. In this case the initial $\chi$ abundance may be much higher or much lower than the equilibrium abundance when it becomes non-relativistic. However, we will see in the next section that the final abundance can still remain very insensitive to initial conditions, where the observed relic abundance is obtained for almost the same parameters.

\section{Phases of Inverse Decay}\label{sec:res}

\begin{figure*}[th!]
    \centering
    \includegraphics[width=1.\columnwidth]{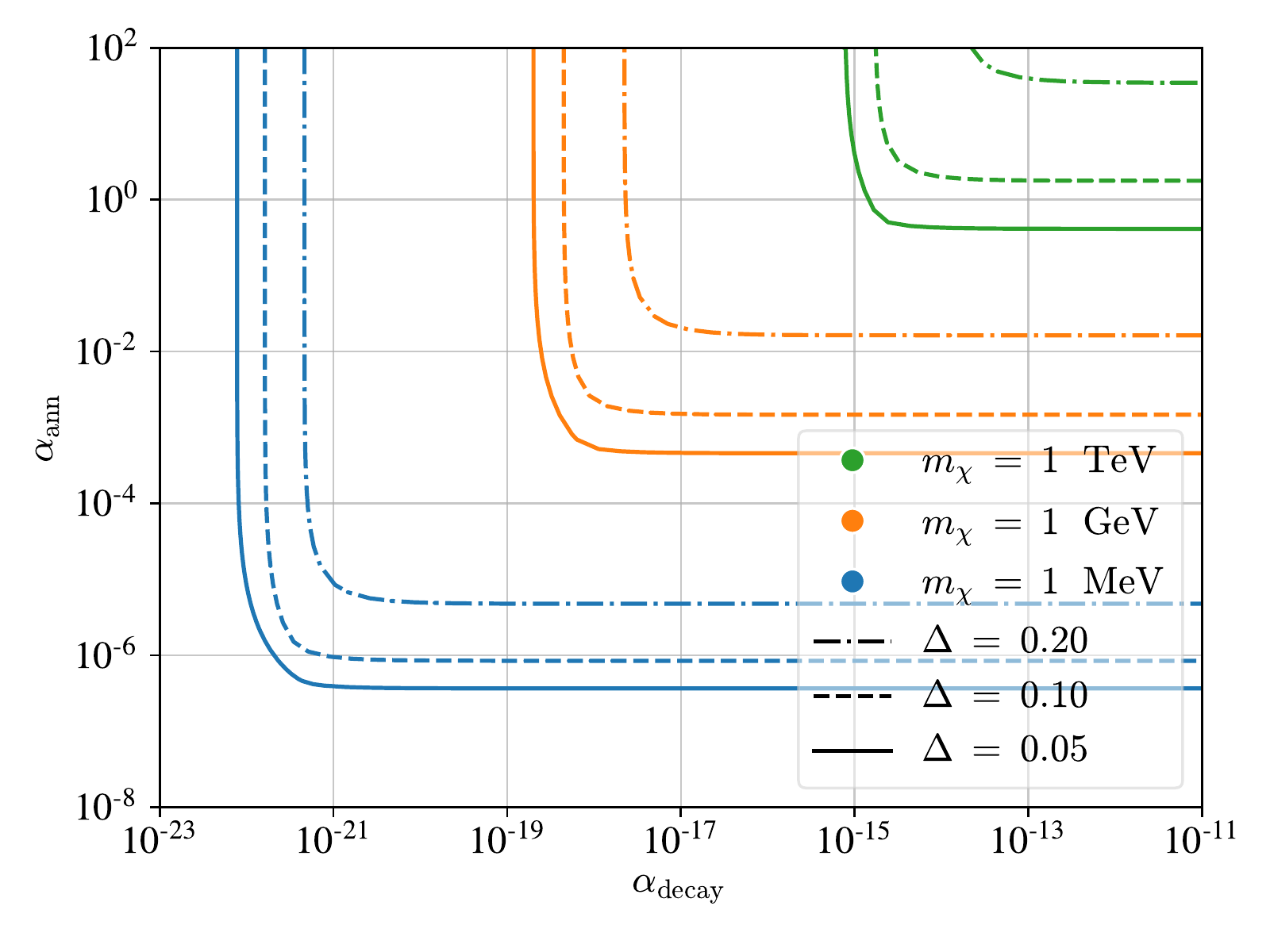}\hfill
    \includegraphics[width=1.\columnwidth]{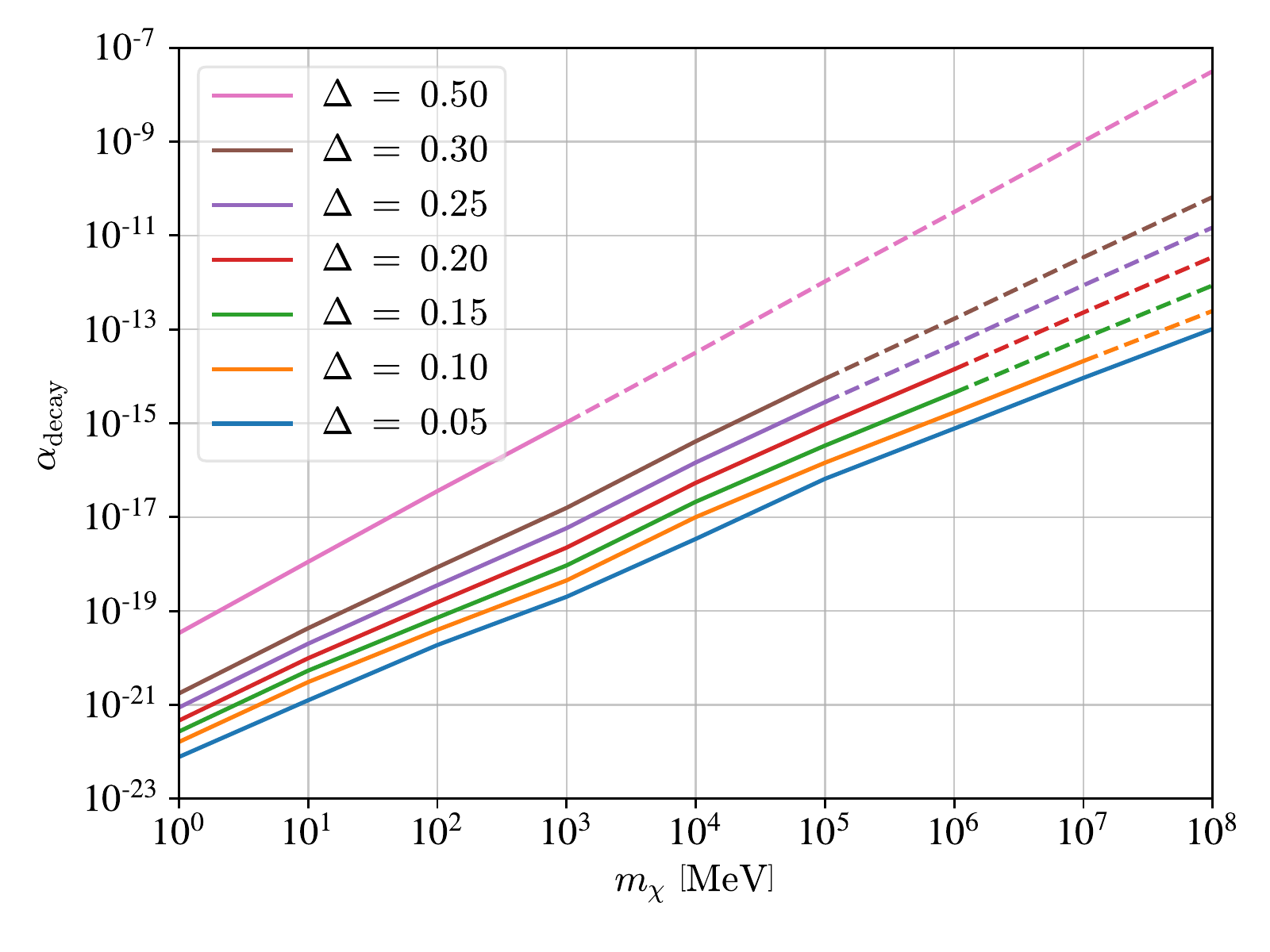}
    \caption{{\bf Left:} 
    Annihilations versus decays phase diagram, for various mass splittings and DM masses (solid colored curves). The two new phases are evident: The first is the horizontal branch,  where the relic abundance of the DM is set through the annihilations of another dark particle, with which it is in chemical equilibrium via decays and inverse decays. The second is the vertical branch, where the relic abundance of the DM is set by the {\it freezeout of the inverse decay} of the DM particle; we dub this `INDY' DM.  {\bf Right:} The decay coupling versus DM mass that reproduces the observed relic abundance along the vertical branch, for various mass splittings. We plot the numerical solution to the Boltzmann equations, with the dashed region of the curves indicating where the effective coupling $\alpha_{\rm ann}$  that we use to parameterize the cross section exceeds $100$.
    }
    \label{fig:phase}
\end{figure*}

We can now setup the system of interest in full. We consider the decay and inverse decay  processes of $\psi \longleftrightarrow\chi \phi$ along with an annihilation process of $\psi \psi \to \tilde \phi \tilde \phi$, with $\phi $ and $\tilde \phi$ indicating a particle in equilibrium with the SM bath---the latter process  maintaining chemical equilibrium between $\psi$ and the SM. The Boltzmann equations for the system are given by:
\beqa \label{eq:BE_chi}
\dot{n}_\chi+3Hn_{\chi}&=&-\langle\Gamma\rangle\left(\frac{n_{\psi}^{\rm eq}}{n_{\chi}^{\rm eq}}n_{\chi}-n_{\psi}\right)\,,\\
\dot{n}_\psi+3Hn_{\psi}&=&\langle\Gamma\rangle\left(\frac{n_{\psi}^{\rm eq}}{n_{\chi}^{\rm eq}}n_{\chi}-n_{\psi}\right)-\langle\sigma v\rangle\left(n_{\psi}^{2}-n_{\psi}^{{\rm eq}^2}\right)\,.\no
\eeqa
Here $\langle \Gamma\rangle$ represents the thermally averaged (time dilation included) decay rate of $\psi \to \chi\phi$, and $\langle \sigma v\rangle$ represents the thermally averaged cross section for the annihilation $\psi \psi \to \tilde \phi \tilde \phi$. We parameterize the cross section as
\beq\label{eq:xsec}
\langle\sigma v\rangle\equiv\frac{\alpha_{\rm ann}^{2}}{m_{\psi}^{2}}\,,
\eeq
and the decay rate as in Eq~\eqref{eq:decay}.

Fig.~\ref{fig:phase} shows our solutions to the Boltzmann equations, $\alpha_{\rm decay}$ versus $\alpha_{\rm ann}$, that reproduces the observed relic abundance of DM, for various mass splitting and  DM masses. There are several distinct phases, each indicating a different mechanism for setting the relic abundance of DM, with some dependence on the initial conditions. Here we focus on two of the phases, assuming that the dark matter is in equilibrium with the bath prior to becoming non-relativistic. 

 Along the horizontal branch, the relic abundance of $\chi$ is being set by the annihilations of $\psi$ into other particles. This is possible because the rapid inverse decays between $\chi$ and $\psi$ keep the two particles in chemical equilibrium. Once $\alpha_{\rm decay}$ is large enough such that $\chi$ and $\psi$ are in chemical equilibrium, its precise value does not matter, and it is $\alpha_{\rm ann}$ that controls the abundance. This branch is similar in spirit to coannihilations~\cite{Griest:1990kh,DAgnolo:2018wcn}, however here the chemical contact between the DM and the particle whose number changing process is determining the relic abundance is set by fast decays and inverse decays, rather than by $2\to2$ processes such as annihilations. 

The value of $\alpha_{\rm ann}$ that reproduces the correct abundance along this branch can be found in the instantaneous freezeout approximation using the same steps as from Eq.~\eqref{eq:xf} to Eq.~\eqref{eq:mvsalpha}, but here $x_f$ is determined by $n_\psi \langle\sigma v\rangle \simeq H$. Doing so, one finds the relationship
\beq
m_\chi \simeq  \alpha_{\rm ann}^{\frac{2}{2+\Delta}}\left( m_{\rm pl} T_{\rm eq}^{1+\Delta} \right)^{\frac{1}{2+\Delta}} \label{eq:mvsann}\,.
\eeq

 In contrast, along the vertical branch, the relic abundance of $\chi$ is being set by the freezeout of inverse decays of the DM. As $\chi$ inverse decays into $\psi$, $\psi$ rapidly annihilated away into other light particles. For large enough $\alpha_{\rm ann}$, its precise value does not play a role, and the relic abundance of $\chi$ is set when its inverse decay shuts off and it  freezes out. This corresponds to the case studied analytically in the previous section, which leads to the DM mass-coupling relationship of Eq.~\eqref{eq:mvsalpha}.

While both branches present a new mechanism for establishing the DM abundance in the early universe, we focus here on the more novel of the two --- the vertical branch, where freezeout of inverse decays of DM set its abundance. We dub this `INverse DecaY' (INDY) dark matter. The parameter space has another two phases, if we assume that initial abundance of $\chi$ is very small: freeze-in, and  freeze-in and freezeout (of inverse decays), which is very similar to the vertical branch. We leave a detailed investigation of the horizontal branch, the other additional phases, and many further details to our companion paper~\cite{future}, and focus here on INDY dark matter.

In  Fig.~\ref{fig:phase} we show $\alpha_{\rm decay}$ as a function of DM mass along the vertical branch, for various mass splittings, where INDY DM is achieved. We show the exact solutions to the Boltzmann equations, in qualitative agreement with the analytic approximations of Eq.~\eqref{eq:mvsalpha}. Note that the minimum value of  $\alpha_{\rm ann}$ needed to be in the INDY phase is given by Eq.~\eqref{eq:mvsann}. 
Therefore the mass of the DM is bounded above by the unitarity of the $\langle\sigma v\rangle$ cross section.
Moreover, since $\Delta>0$, the INDY mechanism predicts lighter DM than the standard WIMP case (which would correspond to $\Delta =0$ in Eq.~\eqref{eq:mvsann}).

\begin{figure*}[t!]
    \centering
   \includegraphics[width=1.\columnwidth]{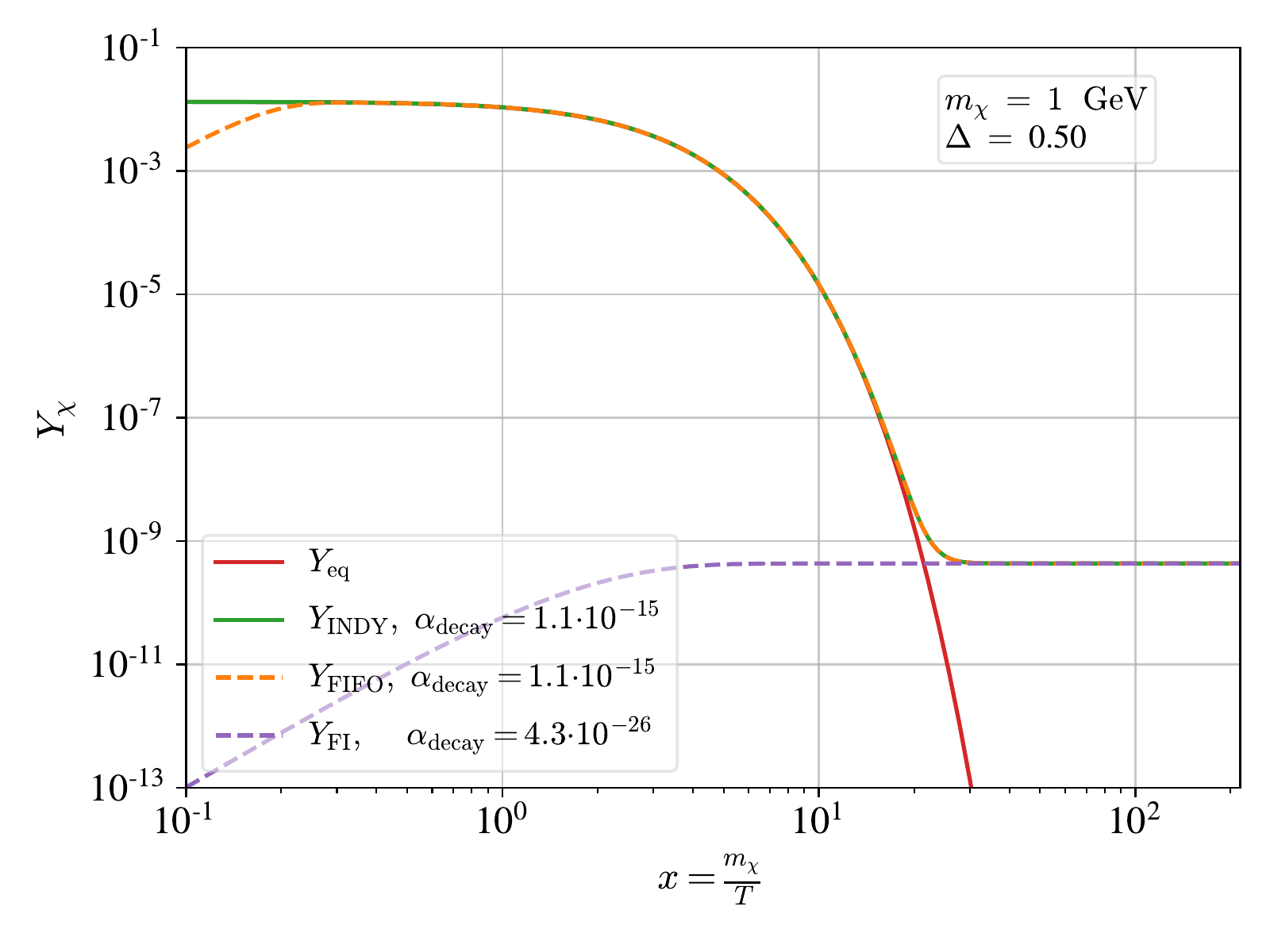}\hfill
   \includegraphics[width=1.\columnwidth]{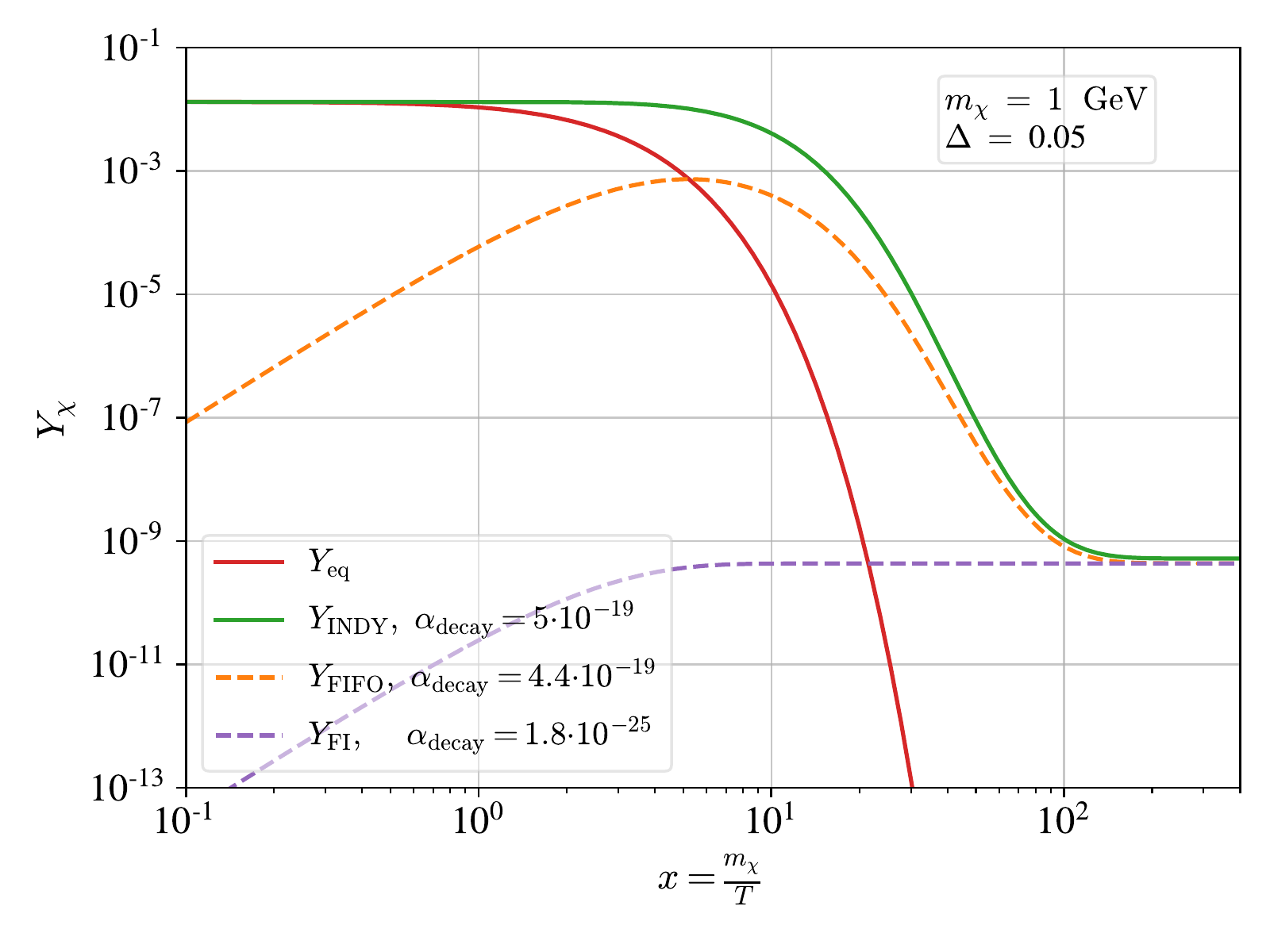}
    \caption{
    Example solutions to the Boltzmann equations along the vertical branch, where the relic abundance of DM is set by the freezeout of the inverse decays of the DM.  For both panels, we show the curves for initial conditions $Y_\chi =Y_\chi^{\rm eq} $ (solid) and $Y_\chi =0$ (dashed), and the corresponding couplings $\alpha_{\rm decay}$ for INDY DM, freeze-in and freezeout ({\tt FIFO}), and freeze-in ({\tt FI}).   {\bf Left:} $\chi$ is thermalized via the inverse decays, and maintains an equilibrium distribution until freezeout of its inverse decays. {\bf Right:} the DM $\chi$ departs from chemical equilibrium early on (due to the smaller $\alpha_{\rm decay}$), and later freezes out.     }
    \label{fig:bench}
\end{figure*}
Fig.~\ref{fig:bench} depicts the evolution of the densities of $\chi$ along the vertical branch of INDYs for some benchmark values of mass, splitting and couplings. The two panels illustrate that the DM density can be set by the freezeout of the inverse decay where $\chi$ either maintains chemical equilibrium with $\psi$ until freezeout occurs  ({\it left panel}), or $\chi$ can decouple at early times and still freezeout in a similar manner ({\it right panel}). In both cases, DM is indeed freezing out non-relativistically, justifying our back-of-the-envelope computation.

A comment is in order regarding initial conditions. In the above we have assumed an equilibrium distribution for the DM $\chi$ at very early times. However in some regions along the vertical branch, where the coupling $\alpha_{\rm decay}$ is very small, decays and inverse decays may not be strong enough to bring $\chi$ into equilibrium in the early universe. To understand the impact of the initial conditions, we assess the extent to which varying them changes the relic abundance.  
We find our results to be insensitive to the initial conditions over a broad range of DM masses and mass splittings, allowing  many orders of magnitude change in the initial condition without significantly modifying the relic abundance. If the dark matter inverse decays thermalize before they become non-relativistic, then the final relic abundance is essentially insensitive to the initial conditions. Additionally, one can consider the case in which the inverse decays are never in equilibrium. In this case, starting with a near equilibrium abundance or even no abundance of dark matter (in which case dark matter freezes in and then freezes out) produces almost the identical parameter space to match the DM relic abundance. This is evident in the right panel of Fig.~\ref{fig:bench}, where relic abundance for starting with an equilibrium value (labeled {\tt INDY}) and starting with a small abundance (labeled {\tt FIFO}) produces nearly identical $\alpha_{\rm decay}$ values. We relegate a detailed discussion of initial conditions and the relation to other phases of the parameter space to our companion paper~\cite{future}.

\section{Model}\label{sec:toy}
 
 As proof of concept, we now present a model that realizes INDY dark matter.  Consider a $U(1)_{d}$ gauge theory with gauge coupling $ e_{d} $,
two Dirac fermions $ \chi $ and $ \psi $, and a complex scalar $ \phi $, with charges $0$, $1$, and $1$ respectively. The general renormalizable Lagrangian contains masses for the fermion, a spontaneous symmetry breaking potential for the scalar, and the Yukawa interaction
 \beq
\mathcal{L}_{\text{Yukawa}} = -  y\phi^{*}\bar{\chi}\psi+h.c .
\eeq
 The field $ \phi $ acquires a VEV through spontaneous symmetry breaking,  which gives a mass $m_{A_{d}}$ to the dark photon  and sources a mixing term between the fermions. This leads to interactions between $ \psi $, $ \chi $ and the dark photon that yield decays and inverse decays. Mapping to the parametrization of Eqs.~\eqref{eq:xsec} and \eqref{eq:decay}, we have \cite{Shtabovenko:2020gxv,Shtabovenko:2016sxi,Mertig:1990an}:
 \beq
\alpha_{\rm decay}=\frac{y^{2}\left(\left(m_{\chi}-m_{\psi}\right)^{2}-m_{A_{d}}^{2}\right)\left(\left(m_{\chi}+m_{\psi}\right)^{2}+2m_{A_{d}}^{2}\right)}{32\pi m_{\psi}^{2}\left(m_{\chi}-m_{\psi}\right)^{2}}
 \eeq
and
 \beq
\alpha_{\rm ann}^{2}=\frac{e_{d}^{4}\sqrt{1-\frac{m_{A_{d}}^{2}}{m_{\psi}^{2}}}\left(m_{\psi}^{2}-m_{A_{d}}^{2}\right)}{32\pi\left(m_{\psi}^{2}-\frac{1}{2}m_{A_{d}}^{2}\right)^{2}}m_{\psi}^{2}\,.
\eeq   
Large values of $ y $ and low values of $ e_d $  correspond to large $ \alpha_{\rm decay} $ and represents solutions on the horizontal branch of Fig. \ref{fig:phase}, while small values of $ y $ and large values of $ e_d $ correspond to large values of $ \alpha_{\rm ann} $ and represents solutions along the vertical branch of INDY dark matter.

For dark matter masses at the MeV scale and above,  other processes such as co-scattering~\cite{DAgnolo:2017dbv} with the dark photon can make small corrections to the curves in Fig.~\ref{fig:phase}. On the horizontal branch, co-scattering plays no role as it does not change the number of $ n_{\chi}+n_{\psi} $ particles. INDY dark matter requires large values of $ \alpha_{\rm ann} $, which correspond to large values of dark coupling $ e_{d} \sim {\cal O}(10^{-2}-1)$, resulting in the dark photon  mass being similar to the mass splitting between $ \chi $ and $ \psi $. Since the ratio between the rates of the co-scattering process and the inverse decay is proportional to the dark photon number density, the rate of the co-scattering processes decreases rapidly when $x\sim\Delta^{-1}\sim m_{\chi}/m_{A_{d}}$, leaving the inverse decay as the dominant process. Co-scattering with the scalar and co-annihilation processes are small for similar reasons, and the annihilation process of $ \chi\chi \rightarrow  A_{d} A_d $ is suppressed by $ y^4 $. The model can thus accommodate the INDY mechanism. As a benchmark, for $ m_{\chi} = 200\  {\rm MeV} $, 
$ m_{\psi} = 230 \ {\rm MeV}$, $ m_{A_d} = 27\  {\rm MeV} $, $ m_\phi = 60 \  {\rm MeV} $, $ e_{d} = 0.32 $ and  $y=7.7\times 10^{-9}$ and $ \epsilon = 
10^{-4} $ we obtain the observed relic abundance of DM particles along the vertical INDY branch, evading current constraints from BBN and cosmology \cite{Parker:2018vye,Banerjee:2019pds,BaBar:2017tiz,Andreas:2012mt,BLUMLEIN2014320,Batley_2015178,Tsai:2020vpi,Ibe_2020,Sabti_2020,Alexander:2016aln,Bauer:2018onh}. Equilibrium with the bath is maintained via dark photon decay and inverse decays, $A_d \leftrightarrow e^+ e^-$.

\begin{figure}[t!]
    \centering
\includegraphics[width=1.\columnwidth]{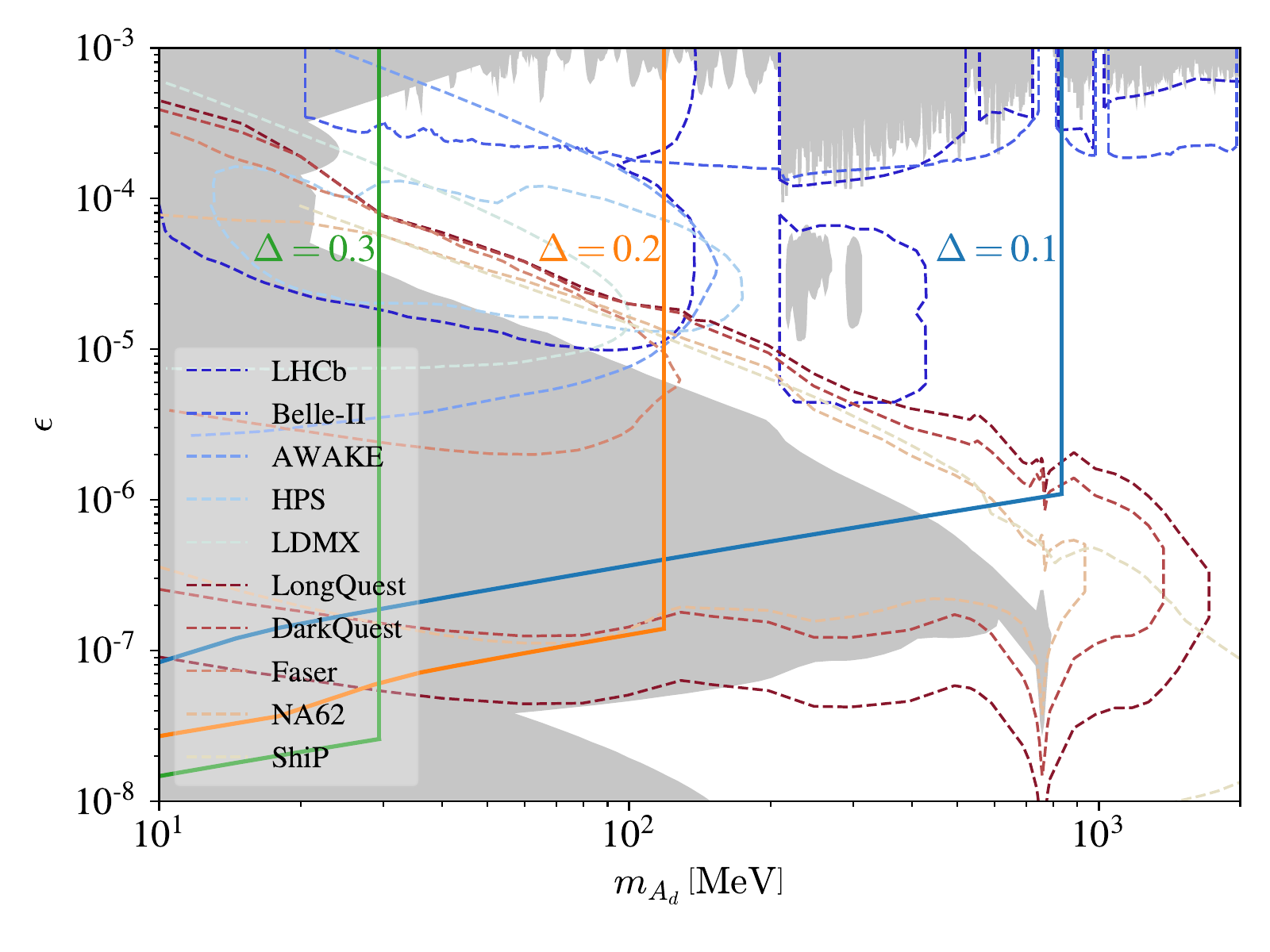}
    \caption{Allowed dark photon parameter space. Solid colored curves indicate the minimal kinetic mixing as a function of dark photon mass within the model, for  various values of mass splitting $\Delta$ and with $m_{A_d} = 0.9\, \Delta m_\chi$. We show existing limits~\cite{Fradette:2014sza,Alexander:2016aln,Chang:2016ntp,Hardy:2016kme,Pospelov:2017kep,NA64:2018lsq,LHCb:2017trq,LHCb:2019vmc,Parker:2018vye,Tsai:2019buq} in shaded gray, with future projections~\cite{Celentano:2014wya,Jaegle:2015fme,Ilten:2015hya,Alekhin:2015byh,Ilten:2016tkc,Alexander:2016aln,Caldwell:2018atq,Berlin:2018pwi,Berlin:2018bsc,FASER:2018eoc,NA62:2312430,Tsai:2019buq} indicated by the colored dashed curves. 
     }
    \label{fig:exp}
\end{figure}

\section{Phenomenology}\label{sec:toy}

At the mechanism level, $\chi$ does not couple directly to the SM. In the model above, all interactions of the $\chi$ particle are  suppressed by $y\sim 10^{-8}$, and the couplings to the SM are  further suppressed by $\epsilon \ll 1$. As a result, direct detection of DM in the laboratory and indirect detection of DM  annihilations in the galaxy will be highly suppressed, and thus INDY DM can be expected to evade these conventional searches. For other examples of exponentially small couplings to the SM from coannihilations, see Ref.~\cite{DAgnolo:2019zkf}.

The dark photon can be searched for directly. 
Fig.~\ref{fig:exp} shows the relevant parameter space for a visible decaying dark photon. Existing limits are indicated in shaded gray~\cite{Fradette:2014sza,Alexander:2016aln,Chang:2016ntp,Hardy:2016kme,Pospelov:2017kep,NA64:2018lsq,LHCb:2017trq,LHCb:2019vmc,Parker:2018vye,Tsai:2019buq}. Solid curves indicate the lower limit on the kinetic mixing parameter $\epsilon$ for given mass splitting $\Delta$, where we have fixed $m_{A_d}=0.9\Delta m_\chi$. Below the solid curves, the SM and dark sector are not thermalized, while to the right of the curve, one does not reproduce the DM relic abundance since the annihilation rate is too small. As is evident, the allowed parameter space is expected to be probed by future experiments, indicated by the dashed curves~\cite{Celentano:2014wya,Jaegle:2015fme,Ilten:2015hya,Alekhin:2015byh,Ilten:2016tkc,Alexander:2016aln,Caldwell:2018atq,Berlin:2018pwi,Berlin:2018bsc,FASER:2018eoc,NA62:2312430,Tsai:2019buq} (bounds and projections are taken from the recent compilation in Ref.~\cite{Tsai:2020vpi}).

Discovery of the INDY dark matter candidate can hopefully be made via the direct production of the $\psi$ particle, which decays into $\chi$ and either additional invisible products or SM products. The $\psi$ particle is typically long lived; for sub-GeV INDY DM, the lifetime can vary from $\mathcal{O}({\rm mm})$ to  $\mathcal{O}(10^{4}~ {\rm km})$. In the model presented above, we have the decay chain $\psi \to \chi  A_d$, $A_d \to e^+ e^-$. This decay chain can be searched for in beam dump experiments where off-shell dark photons can be produced. A similar scenario was recently considered in Ref.~\cite{Tsai:2019buq}, which showed that such decay chains can be probed in current and future beam dump experiments such as NA62~\cite{NA62:2017rwk,Dobrich:2018ezn,Dobrich:2019dxc,Chiang:2016cyf,NA62:2019meo}, and SeaQuest/DarkQuest/LongQuest~\cite{Gardner:2015wea,Berlin:2018pwi,Liu:2017ryd}.

\mysections{Acknowledgments}
  The work of YH is supported by the Israel Science Foundation (grant No. 1112/17), by the Binational Science Foundation (grant No. 2018140), by the I-CORE Program of the Planning Budgeting Committee (grant No. 1937/12), and by the Azrieli Foundation. EK is supported by the Israel Science Foundation (grant No. 1111/17), by the Binational Science Foundation  (grants No. 2016153 and 2020220) and by the I-CORE Program of the Planning Budgeting Committee (grant No. 1937/12).  
The work of HM was supported by the Director, Office of Science, Office of
High Energy Physics of the U.S. Department of Energy under the
Contract No. DE-AC02-05CH11231, by the NSF grant
PHY-1915314, by the Binational Science Foundation (grant No. 2018140), by the JSPS Grant-in-Aid for
Scientific Research JP20K03942, MEXT Grant-in-Aid for Transformative Research Areas (A)
JP20H05850, JP20A203, by WPI, MEXT, Japan, the Institute for AI and Beyond of the University of Tokyo, and  Hamamatsu Photonics, K.K.

\bibliography{biblio}{}

\end{document}